\documentclass[twocolumn,prb,showpacs,amsmath,amssymb,english]{revtex4}
\usepackage{graphicx}
\usepackage{dcolumn}
\usepackage{color}
\begin{document}

\title {Point-like spin-dependent interaction\\ in calculations of self-energy ladder diagrams}

\author{I. A. Nechaev$^1$}
\altaffiliation[Also at: ]{Theoretical Physics Department, Kostroma State University, st. 1-st of May, 14,
156961 Kostroma, Russia.}
\author{I. Nagy$^{2,1}$}
\author{P. M. Echenique$^{1,3}$}
\author{E. V. Chulkov$^{1,3}$}
\affiliation{$^1$Donostia International Physics Center (DIPC), \\P.
Manuel de Lardizabal 4, 20018 San
Sebasti{\'a}n, Spain\\
$^2$Department of Theoretical Physics, Institute of Physics, Technical University of Budapest, H-1521 Budapest, Hungary\\
$^3$Departamento de F{\'\i}sica de Materiales, Facultad de Ciencias
Qu{\'\i}micas, UPV/EHU and Centro Mixto CSIC-UPV/EHU, Apdo. 1072,
20080 San Sebasti\'an, Spain}

\date{\today}

\begin{abstract}
An instantaneous and zero-range spin-dependent interaction, derived by summing an infinite number of
electron-hole ladder diagrams within a local approximation, is analyzed as a function of the electron gas
density and the relative spin polarization. The strength of such an interaction is defined by an integral of
a statically screened Coulomb interaction with a spatially localized weighting factor. This weighting factor
represents the mutually uncorrelated motion of an electron-hole pair in singlet or triplet spin states
($S^z=0,\pm1$). An implementation, based on a Yukawa-type interaction with a spin-polarization-dependent
Thomas-Fermi screening length, is given.

\end{abstract}


\maketitle

A rigorous study, based on quantum many-body perturbation theory\cite{FettWal} as applied to excited
quasiparticle states in itinerant ferromagnetic materials, requires an approach to the quasiparticle
self-energy accounting for the contribution of both charge- and spin-density fluctuations.\cite{Itiner}
Within this approach, due to the correspondence between multiple electron-hole ($e-h$) scattering events and
a spin fluctuation,\cite{SFl,PeCar} the self-energy can be treated as an integral over the four-point $e-h$
scattering amplitude.\cite{Bickers_book,GV_electron_liquid}

The theoretical problem, which appears already at the homogeneous electron gas level, lies in this amplitude
and can be largely simplified in the following ways: (i) by using {\it a point-like interaction} with
strength defined by the Hubbard $U$ parameter;\cite{Hub_U} (ii) by introducing {\it a local pseudopotential}
obtained within some self-consistently closed procedure;\cite{Bickers_book} (iii) by reducing an infinite set
of diagrams to {\it a local-field factor} \cite{GV_electron_liquid} tabulated and parameterized by making use
of accurate quantum Monte Carlo data for the electron gas.

Apart from the obvious theoretical interest for a prototype homogeneous system, real materials give an
additional important challenge. In evaluating the quantities listed above, one should properly take into
consideration the band-structure effects which surely influence calculated characteristics of excited
quasiparticle states. However, the obtained results are very sensitive to an approximated form chosen for
these quantities.\cite{Diff_App_LFF,U_calcs} Therefore, the question of how in {\it ab initio} calculations
to sum an infinite number of diagrams, which contribute to the four-point $e-h$ scattering amplitude, is
still open.

The standard formulation\cite{Hedin} for excited states starts by defining a dynamical,
\emph{spin-independent} dielectric screening of the bare Coulomb interaction between two test charges. This
concept of a polarizable medium allows a straightforward and successful implementation of the
Hartree-Fock-like technique with such a screened interaction instead of the bare one (the $GW$ method), since
it is based on a complete set of single-particle states at the Hartree (mean-field) level. Thus the required
approach naturally should start with the above screened interaction to consider multiple scattering effects
in the important $e-h$ channel.

Here we present a theoretical study which rests on Hedin's concept.\cite{Hedin} We derive a \emph{point-like}
\emph{spin-dependent} interaction. The foremost advantage of using the latter is that a number of many-body
problems becomes more easily solvable. To achieve this goal we shall use the approach of Ref.
\onlinecite{IAN05} which was formulated for uniform systems in momentum space. This approach is based on a
treatment of the self-energy ladder diagrams and does not contain adjustable parameters or quantities defined
outside the scope of the approach. A key quantity of the approach is a local interaction
$\widetilde{W}_{\sigma \sigma'}$ that (i) ensures a simplification similar to Hubbard models, (ii) includes
charge and spin fluctuations, and (iii) has a transparent connection with the exchange part of the many-body
local field factor.

Accomplishing our goal, first, we perform a real-space analysis of the approach of Ref.~\onlinecite{IAN05}.
Such an analysis allows one to derive basic equations of the approach to be applied to a broader spectrum of
materials. Unless stated otherwise, atomic units, $e^2=\hbar=m=1$, are used throughout this work. By applying
standard notation for the coordinate, $\bar{1}\equiv(\sigma_1,\mathbf{r}_1,t_1)$, in which the quasiparticle
spin is denoted as $\sigma_1$, the Bethe-Salpeter equation defining the $e-h$ scattering amplitude (the $T$
matrix) can be written as\cite{Bickers_book}

\begin{eqnarray}\label{complete_e_h_vertex}
T(\bar{1},\bar{2}|\bar{3},\bar{4})&=&T^{0}(\bar{1},\bar{2}|\bar{1'},\bar{2'})\left[ \delta(\bar{1'}-\bar{3})
\delta(\bar{2'}-\bar{4})\right. \nonumber\\
&+&\left.K^{0}(\bar{1'},\bar{2'}|\bar{3'},\bar{4'})T(\bar{3'},\bar{4'}|\bar{3},\bar{4})\right], 
\end{eqnarray}
where $T^{0}(\bar{1},\bar{2}|\bar{1'},\bar{2'})$ is the irreducible $e-h$ vertex and the free propagator for
$e-h$ pair is given by the product of the Green functions as 
$K^{0}(\bar{1},\bar{2}|\bar{1'},\bar{2'})=iG(\bar{1},\bar{1'})G(\bar{2'},\bar{2})$.
In Eq.~(\ref{complete_e_h_vertex}), an integral over space-time coordinates and a sum over spin are implied
for repeated variables.

To perform a selective summation of ladder diagrams\cite{IAN05} along the line of Hedin's
expansion\cite{Hedin} for the self-energy in terms of a dynamically screened Coulomb interaction $W$, the
irreducible $e-h$ vertex can be chosen as
\begin{equation*}\label{T_irr_W}
T^{0}(\bar{1},\bar{2}|\bar{3},\bar{4}) = W(1,2) \delta(1-3) \delta(2-4) \delta_{\sigma_1\sigma_3}
\delta_{\sigma_2\sigma_4},
\end{equation*}
where in the right-hand side the spin variables are explicitly written out and space-time coordinates are
abbreviated as $1\equiv(\mathbf{r}_1,t_1)$. As a result, the $T$ matrix is defined by the ladder
approximation to the Bethe-Salpeter equation as
\begin{eqnarray}\label{RSTtype}
    T_{\sigma \sigma '}(1,2|3,4)&=&W(1,2)\left[\delta(1-3)\delta(2-4)\right.\nonumber\\
    &+&\int\,d1'd2'K_{\sigma \sigma'}^{0}(1,2|1',2')\nonumber\\
    &\times& \left.T_{\sigma \sigma '}(1',2'|3,4)\right],
\end{eqnarray}
where the $e-h$ propagator expressed in terms of the Green functions diagonal in spin space. In
Eq.~(\ref{RSTtype}), there is no sum over spin implied and all integrals are written explicitly. The $T$
matrix (\ref{RSTtype}) describes propagation of a mutually correlated $e-h$ pair carrying spin
$S^z=0,\pm1$.\cite{spin_pr} This correlation is realized by repeatedly interacting pair-constituents via
$W(1,2)$ in the intermediate state.

In real space, the approximation done in Ref. \onlinecite{IAN05} results in the $T$ matrix given by
\begin{equation}\label{T_Gamma_delta} T_{\sigma \sigma'}(1,2|3,4)=\tilde{\Gamma}_{\sigma
\sigma'}(1|4)\delta(1-2)\delta(3-4),
\end{equation}
where
\begin{eqnarray}\label{T_matrix_RS}
\tilde{\Gamma}_{\sigma \sigma'}(1|4)&=&\widetilde{W}_{\sigma \sigma'}(1,4)+\int d1'd2' \widetilde{W}_{\sigma
\sigma'}(1,1') \nonumber \\ &\times& K_{\sigma \sigma'}^{0}(1',1'|2',2')\tilde{\Gamma}_{\sigma
\sigma'}(2'|4).
\end{eqnarray}
The interaction $\widetilde{W}_{\sigma \sigma'}(1,4)$ is defined by the equation
\begin{eqnarray}\label{LI_RS}
\int d1'd2' K_{\sigma \sigma'}^{0}(1,1|1',1') \widetilde{W}_{\sigma \sigma'}(1',2') K_{\sigma
\sigma'}^{0}(2',2'|4,4) \nonumber \\ = \int d1' d2' K_{\sigma \sigma'}^{0}(1,1|1',2') W(1',2') K_{\sigma
\sigma'}^{0}(1',2'|4,4).
\end{eqnarray}
The obtained $T$ matrix is a \emph{local} one describing scattering processes in which the coordinates of the
$e-h$ pair both for initial states and for final states coincide. This locality essentially simplifies the
self-energy evaluation as an integral over the $T$ matrix. Actually, owing to Eqs. (\ref{T_Gamma_delta}) and
(\ref{T_matrix_RS}), the $T$-matrix contribution as an additional term to the $GW$ self-energy
$\Sigma^{GW}_{\sigma}(1,2)=iG_{\sigma}(1,2)W(1,2)$ can be cast into the $GW$-like form given by
\begin{equation}\label{STM}
\Sigma_{\sigma}^{T}(1,2)=-i\sum_{\sigma'}G_{\sigma'}(1,2){\cal T}_{\sigma\sigma'}(1|2),
\end{equation}
where
\begin{eqnarray}\label{T_3o}
{\cal T}_{\sigma\sigma'}(1|2)&=&\int d1'd2' \widetilde{W}_{\sigma \sigma'}(1,1') K_{\sigma
\sigma'}^{0}(1',1'|2',2')\nonumber \\ &\times&\left[\tilde{\Gamma}_{\sigma
\sigma'}(2'|2)-\widetilde{W}_{\sigma \sigma'}(2',2)\right].
\end{eqnarray}
Thus, in real space we have obtained the $T$-matrix contribution by exploiting the local approximation which
seems to be reasonable for the corrections to the $GW$ term (see, e.g., Ref.~\onlinecite{Zein06}).

Now we can address to our main goal, deriving an instantaneous and zero-range potential
\begin{equation}\label{point_inter}
\widetilde{W}_{\sigma \sigma'}(1,2)=V_{\sigma \sigma'}^0\delta(1-2).
\end{equation}
It is obvious that in this case 
the irreducible $e-h$ vertex becomes highly local as in Hubbard models. 
Further, we will be guided by the results of Ref.~\onlinecite{IAN06}, where for a uniform system and at small
four-momentum transfer along the $e-h$ channel the local interaction $\widetilde{W}_{\sigma \sigma'}$ was
obtained to be constant in momentum space. Consequently, its real space equivalent corresponds to a
$\delta$-function interaction we need: the resulting $T$ matrix has the form which can be obtained from Eq.
(\ref{RSTtype}) by using the \emph{zero-range} \emph{spin-dependent} interaction (\ref{point_inter}) instead
of $W(1,2)$.

In terms of the noninteracting $e-h$ Green function
\begin{eqnarray*}\label{fehG}
\mathcal{G}^{0}_{\mathbf{k}\sigma\sigma'}(\mathbf{q},\omega) &=&
\frac{\left[1-n_F(\epsilon_{\mathbf{k}\sigma'})\right]
n_F(\epsilon_{\mathbf{k}+\mathbf{q}\sigma})}{\omega-\epsilon_{\mathbf{k}+\mathbf{q}\sigma} +
\epsilon_{\mathbf{k}\sigma'}-i\eta} \nonumber \\
&-&\frac{n_F(\epsilon_{\mathbf{k}\sigma'})\left[1-n_F(\epsilon_{\mathbf{k}+\mathbf{q}\sigma})\right]
}{\omega-\epsilon_{\mathbf{k}+\mathbf{q}\sigma} + \epsilon_{\mathbf{k}\sigma'}+i\eta},
\end{eqnarray*}
where $n_F$ is the Fermi distribution function and the energy $\epsilon_{\mathbf{k}\sigma}$ is measured from
the Fermi energy $\epsilon_F$, the free $e-h$ propagator is given by
\begin{equation*}\label{K_uniform}
K_{\sigma\sigma'}^{0}(\mathbf{q},\omega)=\int \frac{d\mathbf{k}}{(2\pi)^{3}}
\,\mathcal{G}^{0}_{\mathbf{k}\sigma\sigma'}(\mathbf{q},\omega).
\end{equation*}
We introduce a quantity which characterize the mutually uncorrelated motion of the $e-h$ pair as
\begin{equation}\label{xi_general}
\xi_{\sigma\sigma'}(\mathbf{r}) = \frac{1}{K_{\sigma\sigma'}^{0}(0,0)} \int \frac{d\mathbf{k}}{(2\pi)^{3}} \,
e^{i\mathbf{kr}} \lim_{\mathbf{q}\rightarrow0}\mathcal{G}^{0}_{\mathbf{k}\sigma\sigma'}(\mathbf{q},0).
\end{equation}
This quantity is unity at $r=0$ and tends to zero as $\sim1/r$ at $r\rightarrow\infty$. Thus, we can write
the point-like interaction strength in real space as
\begin{equation}\label{pot_str}
V_{\sigma \sigma'}^0=\int d\mathbf{r}|\xi_{\sigma\sigma'}(\mathbf{r})|^2 V(\mathbf{r}),
\end{equation}
with a statically screened Coulomb interaction denoted as $V(\mathbf{r})\equiv W(\mathbf{r},\omega=0)$. In
this expression, $|\xi_{\sigma\sigma'}(\mathbf{r})|^2$ can be considered as a spatially localized weighting
factor.

In models based on the underlying Hartree approach, the interaction in Eq.~(\ref{pot_str}) can be
approximated by a simple Yukawa form $(1/r)\exp(-\lambda r)$, where $\lambda$ specifies a characteristic
length of the screening in the system [e.g., in Thomas-Fermi approximation
$\lambda\equiv\lambda_{TF}=(4k_F/\pi)^{1/2}$]. For completeness, we have to note that in Hartree-Fock-type
approximations for electron-electron interactions the screening length
can\cite{Engel97,Nagy03,Corona04,Nagy04} differ (${\lambda}\sim{k_F}$) from the Thomas-Fermi one. Hedin's
expansion, being our basic frame in the present work, rests on $W(1,2)$ which corresponds to the dielectric
(Hartree) screening between two test charges.

In order to highlight the physical meaning of the quantities $\xi_{\sigma\sigma'}(\mathbf{r})$ and $V_{\sigma
\sigma'}^0$ entering Eq. (\ref{pot_str}), first, we consider the homogenous electron gas (HEG) in the
paramagnetic state. In this case $\xi_{\sigma\sigma'}(\mathbf{r})$ [denoted as $\xi_{p}(\mathbf{r})$] can be
easily found as $\xi_{p}(\mathbf{r})=j_0(k_Fr)$. Here $j_l$ is the spherical Bessel function of order $l$ and
$k_F$ is the Fermi wave vector. Thus, the point-like interaction of strength $V^0_{p}$ can be identified with
the Fermi-Huang pseudopotential\cite{FH_ps} $-2\pi (\tan\delta_0/\mu k)\delta(\mathbf{r})$ at $k=k_F$ ($\mu$
is the reduced mass of the interacting pair) with the phase shift $\delta_0$ evaluated within the first-order
Born approximation as $\tan\delta_0\approx-(\mu k/2\pi)\int d\mathbf{r} [j_0(kr)]^2V(\mathbf{r})$.

As a digression, we would like to notice here that a somewhat similar idea was applied\cite{Bedell87} earlier
to the Landau theory of liquid $^{3}$He. In Ref.~\onlinecite{Bedell87}, the quasiparticle scattering
amplitude was described by the Fourier transform $\int
e^{i\mathbf{q}\mathbf{r}}V_{eff}(\mathbf{r},k)d\mathbf{r}$ of the effective potential
$V_{eff}(\mathbf{r},k)=V_{0}(r)+ (1/3)k^2r^2V_{2}(r)$, where $V_{0}$ and $V_{2}$ are some local potentials.
This effective-potential form addressed the question of non-locality at a given scattering momentum $k$. In
the forward direction ($\mathbf{q}=0$) and $k=k_F$, this amplitude is formally very close to the interaction
strength in question, because for small $k_F r$ one gets $[j_0(k_Fr)]^2\approx1-(1/3)k_F^2 r^2$.

\begin{figure}[tbp] \centering
 \includegraphics[angle=-90,scale=0.5]{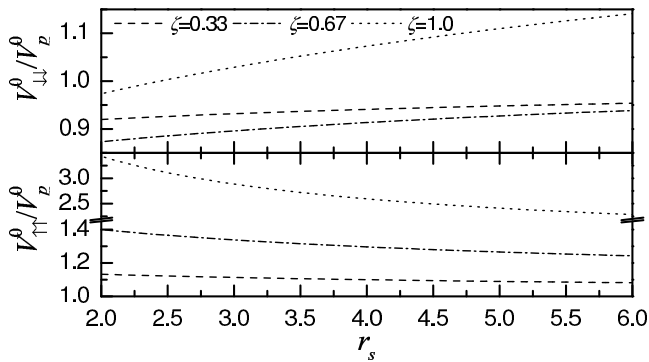}
\caption{The spin-diagonal part of the point-like interaction strength $V_{\sigma\sigma}^0$ over its value in
the paramagnetic state as a function of $r_s$ at the SP $\zeta=0.33$, $0.67$, and $1.0$.}\label{v_diag_rs}
\end{figure}
\begin{figure}[tbp]
\centering
 \includegraphics[angle=-90,scale=0.5]{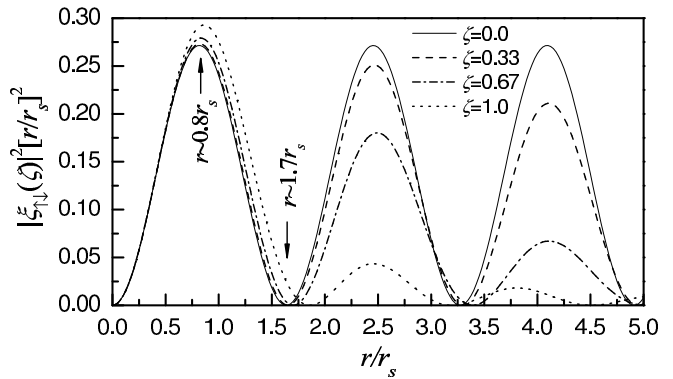}
\caption{The spin-non-diagonal part of $\xi_{\sigma\sigma'}$ as a function of $r$ at different values of the
SP $\zeta$.}\label{xi_non_diag_r}
\end{figure}
\begin{figure}[tbp]
\centering
 \includegraphics[angle=-90,scale=0.5]{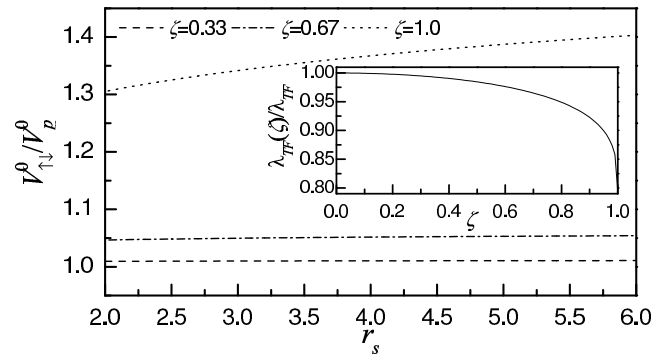}
\caption{The spin-non-diagonal part of the point-like interaction strength $V_{\uparrow\downarrow}^0$ over
its value in the paramagnetic state as a function of $r_s$ at the SP $\zeta=0.33$, $0.67$, and $1.0$. Inset:
the ratio $\lambda_{TF}(\zeta)/\lambda_{TF}$ as a function of $\zeta$.}\label{v_n_diag_rs}
\end{figure}

Using the Yukawa form for $V(\mathbf{r})$, one obtains the following expression for the strength
\begin{equation}\label{V_Yukawa_param}
V_{p}^0=\frac{\pi}{k_F^2}\ln\left(1+4k_F^2/\lambda^2_{TF}\right),
\end{equation}
that in a week scattering regime reproduces the well-known expression for the interaction energy\cite{PeCar}
of the quasiparticles both of which are on the Fermi surface. This $V_{p}^0$ is inversely proportional to
$k_F^2$, showing an expected behavior since the scattering has vanishing effect by growing density.

For a spin-polarized HEG, when the electron and the hole belong to the same $\sigma$ subsystem ($S^z=0$),
$\xi_{\sigma\sigma}(\mathbf{r})=j_0(k_F^{\sigma}r)$, where the Fermi wave vector $k_F^{\sigma}$ for spin
$\sigma$ is expressed as\cite{Moriya} $k_F^{\sigma}=k_F(1-\sigma\zeta)^{1/3}$ with the paramagnetic value of
$k_F$. Here $\sigma=+1$ for spin-up ($\uparrow$) and $\sigma=-1$ for spin-down ($\downarrow$), respectively.
The relative spin polarization (SP) $\zeta$ is given by $\zeta=|n_{\uparrow}-n_{\downarrow}|/n$, where
$n_{\sigma}$ is the spin $\sigma$ electron density, $n=n_{\uparrow}+n_{\downarrow}$ being the total electron
density.

In addition to $\xi_{\sigma\sigma}(\mathbf{r})$, the $\zeta$ dependence of the strength $V_{\sigma\sigma}^0$
comes from $\lambda_{TF}(\zeta)={\lambda}_{TF}\sqrt{\frac{1}{2}\sum_{\sigma}(1-\sigma\zeta)^{1/3}}$ obtained
at the long-wave limit from the standard expression for the irreducible
polarizability\cite{GV_electron_liquid,Sarma05} in the spin-polarized case. As a net result, $V_{\sigma
\sigma}^0$ has the same form as Eq.~(\ref{V_Yukawa_param}) but with $k_F^{\sigma}$ and $\lambda_{TF}(\zeta)$
instead of $k_F$ and $\lambda_{TF}$, respectively.

Fig.~\ref{v_diag_rs} illustrates the spin-diagonal part of the interaction strength plotted as the ratio
$V_{\sigma\sigma}^0/V_{p}^0$ for $r_s$ ranging from 2 to 6 (the metallic density range) at different
${\zeta}$ parameters. Due to the nontrivial $\zeta$ dependence resulting from
$\xi_{\downarrow\downarrow}(\mathbf{r})$ and $V(\mathbf{r})$ with $\lambda_{TF}(\zeta)$ in the basic
Eq.~(\ref{pot_str}), as the SP increases $V_{\downarrow\downarrow}^0$ becomes greater or smaller than its
value in the paramagnetic state at fixed $r_s$. On the other hand, the results for
$V_{\uparrow\uparrow}^0/V_{p}^0$ show that at fixed $r_s$ the ratio demonstrates a monotonic increase with
increasing $\zeta$. This increase becomes smaller as $r_s$ increases. However, at $\zeta\rightarrow1$ limit
this monotonic behavior together with decreasing $n_{\uparrow}$ can lead to a nontrivial $\zeta$ dependence
of the corresponding contribution to the quasiparticle self-energy.

The spin-non-diagonal part of $\xi_{\sigma\sigma'}(\mathbf{r})$ corresponds to the case when the electron and
the hole belong to different spin subsystems ($S^z=\pm1$). This part is given by
\begin{equation}\label{non_diagonal}
\xi_{\uparrow\downarrow}(\mathbf{r})=\frac{1}{2}\sum_{\sigma}
\left(1-\frac{\sigma}{\zeta}\right)[j_0(k_F^{\sigma}r)+j_2(k_F^{\sigma}r)].
\end{equation}
Note that $\xi_{\uparrow\downarrow}(\mathbf{r}) = \xi_{\downarrow\uparrow}(\mathbf{r})$ and at $\zeta=0$,
$\xi_{\uparrow\downarrow}(\mathbf{r}) = \xi_{p}(\mathbf{r})$. At the $\zeta\to1$ limit,
$\xi_{\uparrow\downarrow}(\mathbf{r})$ tends to $j_0(k_F^{\downarrow}r)+j_2(k_F^{\downarrow}r)$.

Fig.~\ref{xi_non_diag_r} gives insight into the detailed (radial) behavior of the spin-non-diagonal part of
the weighting factor appearing in Eq.~(\ref{pot_str}). Multiplied by $[r/r_s]^2$, this factor has the first
maximum at $r\sim 0.8 r_s$ ($0.86r_s$ for $\zeta=1$) and the first node at $r\sim 1.7r_s$ ($1.86r_s$ for
$\zeta=1$). Practically, owing to the $r$-dependence of the applied Yukawa interaction, one can focus on the
part of $\xi_{\uparrow\downarrow}(\mathbf{r})$ which corresponds to the $r$-range from zero to the mentioned
node. Within this interval the considered function shows the {\it weak} $\zeta$-dependence. As a result, the
behavior of $V_{\uparrow\downarrow}^0$ plotted in Fig.~\ref{v_n_diag_rs} is mainly caused by the
$\zeta$-dependence of the screening length $\lambda_{TF}(\zeta)$ (see the inset of the figure). As is evident
from the figure, for $\zeta$ up to $0.67$ the shown ratio $V_{\uparrow\downarrow}^0/V_{p}^0$ changes within
$\sim5\%$. This means that Hubbard's idea on constancy is closely satisfied.

A comparison of Figs.~\ref{v_diag_rs} and \ref{v_n_diag_rs} unambiguously shows that it is
$V^{0}_{\uparrow\uparrow}$ in the spin-diagonal part of the point-like interaction strength which exhibits
the most strong changes as a function of the SP in the itinerant many-body system.

In conclusion, we have determined the real-space $e-h$ scattering amplitude and its contribution to the
quasiparticle self-energy within a local approximation by using, as a background, a variational
momentum-space approach.\cite{IAN05} We have shown that for low-energy quasiparticle excitations this
amplitude can be expressed in terms of an instantaneous and zero-range spin-dependent pseudopotential. The
strength $V_{\sigma\sigma'}^0$ of such a pseudopotential is defined by the volume integral of the statically
screened Coulomb interaction with a weighting factor concerned with the mutually uncorrelated motion of the
$e-h$ pair carrying spin $S^z=0,\pm1$. The analysis carried out for the strength shows that the obtained
results, due to the transparent dependencies on those physical variables which are based on Hedin's
polarization concept, could be useful in attempts to go beyond the $GW$ method in an {\it ab initio} study on
itinerant ferromagnets.

We thank V.V. Tugushev for a critical reading of the manuscript and useful discussions. This work was
partially supported by Departamento de Educaci\'on del Gobierno Vasco and MCyT (Grant No. FIS
2004-06490-C03-01). The work of I.N. was supported partly by the Hungarian OTKA (Grant Nos. T046868 and
T049571)



\begin{thebibliography}{00}

\bibitem{FettWal} A.L. Fetter and J.D. Walecka, \emph{Quantum Theory of Many-particle Systems}
(McGraw-Hill, New York, 1971).

\bibitem{Itiner} See, e.g.,
R. Knorren, K.H. Bennemann, R. Burgermeister, and M. Aeschlimann, Phys. Rev. B {\bf61}, 9427 (2000); V.P.
Zhukov, E.V. Chulkov, and P.M. Echenique, Phys. Rev. Lett. {\bf93}, 096401 (2004); J. Sch\"{a}fer, M.
Hoinkis, E. Rotenberg, P. Blaha, and R. Claessen, Phys. Rev. B {\bf72}, 155115 (2005); M. Cinchetti, M.
S\'{a}nchez Albaneda, D. Hoffmann, T. Roth, J.-P. W\"{u}stenberg, M. Krau{\ss}, O. Andreyev, H.C. Schneider,
M. Bauer, and M. Aeschlimann, cond-mat/0605272 (unpublished).

\bibitem{SFl} S. Doniach and S. Engelsberg, Phys. Rev. Lett. {\bf17}, 750 (1966); W.F. Brinkman and S. Engelsberg, Phys. Rev. {\bf169}, 417
(1968).

\bibitem{PeCar} C.J. Pethick and G.M. Carneiro, Phys. Rev. A {\bf7}, 304 (1973).

\bibitem{Bickers_book} N.E. Bickers, in \emph{Theoretical Methods for Strongly Correlated Electrons}, edited
by D. S\'{e}n\'{e}chal, A.-M. Tremblay, and C. Bourbonnais (Springer, New York, 2004).

\bibitem{GV_electron_liquid} G.F. Giuliani and G. Vignale, \emph{Quantum Theory of the Electron Liquid}
(Cambridge University Press, Cambridge, England, 2005).

\bibitem{Hub_U} A. Liebsch, Phys. Rev. B {\bf23}, 5203
(1981); F. Manghi, V. Bellini, and C. Arcangeli, Phys. Rev. B {\bf56}, 7149 (1997).

\bibitem{Diff_App_LFF} See, e.g., F. Sottile, V. Olevano, and L. Reining, Phys. Rev. Lett. {\bf91}, 056402
(2003); G. Adragna, R. Del Sole, and A. Marini, Phys. Rev. B {\bf68}, 165108 (2003); S. Botti, F. Sottile, N.
Vast, V. Olevano, L. Reining, H.-C. Weissker, A. Rubio, G. Onida, R. Del Sole, R. W. Godby, Phys. Rev. B
{\bf69}, 155112 (2004).

\bibitem{U_calcs} See, e.g., F. Aryasetiawan, M. Imada, A. Georges, G. Kotliar, S. Biermann, and A.I. Lichtenstein, Phys.
Rev. B {\bf70}, 195104 (2004); I.V. Solovyev and M. Imada, \textit{ibid.} {\bf71}, 045103 (2005); I.V.
Solovyev, \textit{ibid.} {\bf73}, 155117 (2006); K. Nakamura, R. Arita, Y. Yoshimoto, and S. Tsuneyuki,
cond-mat/0510425 (unpublished); F. Aryasetiawan, K. Karlsson, O. Jepsen, and U. Sch\"{o}nberger,
cond-mat/0603138 (unpublished).

\bibitem{Hedin} L. Hedin, Phys. Rev. {\bf139}, A796 (1965).

\bibitem{IAN05} I.A. Nechaev and E.V. Chulkov, Phys. Rev. B {\bf71}, 115104 (2005).

\bibitem{spin_pr}Here $S^z$ is an extra spin projection caused by exciting the $e-h$ pair in the system and assigned to this
pair.

\bibitem{Zein06} N.E. Zein, S.Y. Savrasov, and G. Kotliar, Phys. Rev. Lett. {\bf96}, 226403 (2006).

\bibitem{IAN06} I.A. Nechaev and E.V. Chulkov, Phys. Rev. B {\bf73}, 165112 (2006).

\bibitem{Engel97} G.E. Engel, Phys. Rev. Lett. {\bf78}, 3515 (1997).

\bibitem{Nagy03} I. Nagy, J.I. Juaristi, R. D\'{i}ez Mui\~{n}o, and P.M. Echenique, Phys. Rev. B {\bf67},
073102 (2003).

\bibitem{Corona04} M. Corona, P. Gori-Giorgi, and J.P. Perdew,
Phys. Rev. B {\bf 69}, 045108 (2004).

\bibitem{Nagy04} I. Nagy, R. D\'{i}ez Mui\~{n}o, J.I. Juaristi, and P.M. Echenique, Phys. Rev. B {\bf69},
233105 (2004).

\bibitem{FH_ps} K. Huang and C.N. Yang, Phys. Rev. {\bf105}, 767 (1957);
A. Derevianko, Phys. Rev. A {\bf72}, 044701 (2005); Z. Idziaszek and T. Calarco, Phys. Rev. Lett. {\bf96},
013201 (2006).

\bibitem{Bedell87} T.L. Ainsworth and K.S. Bedell, Phys. Rev. B {\bf35}, 8425 (1987).

\bibitem{Moriya} T. Moriya, \emph{Spin Fluctuations in Itinerant Electron Magnetism}
(Springer, Berlin, 1985).

\bibitem{Sarma05} Y. Zhang and S. Das Sarma, Phys. Rev. Lett. {\bf95}, 256603 (2005).

\end{thebibliography}
\end{document}